\documentclass[a4paper, amsfonts, amssymb, amsmath, reprint, showkeys, nofootinbib, twoside]{revtex4-1}
\usepackage[english]{babel}
\usepackage[utf8]{inputenc}
\usepackage[colorinlistoftodos, color=green!40, prependcaption]{todonotes}
\usepackage{amsthm}
\usepackage{mathtools}
\usepackage{physics}
\usepackage{xcolor}
\usepackage{graphicx}
\usepackage[left=23mm,right=13mm,top=35mm,columnsep=15pt]{geometry} 
\usepackage{adjustbox}
\usepackage{placeins}
\usepackage[T1]{fontenc}
\usepackage{lipsum}
\usepackage{csquotes}
\usepackage[pdftex, pdftitle={Article}, pdfauthor={Author}]{hyperref} 

\begin{document}
\title{First-principles calculation of the electronic and optical properties of Gd$_2$FeCrO$_6$ double perovskite: Effect of Hubbard U parameter}

\author{Subrata Das \textsuperscript{a,b}, M. D. I. Bhuyan \textsuperscript{a} and M. A. Basith \textsuperscript{a}}
    \email[Email address: ]{mabasith@phy.buet.ac.bd}
    \affiliation{\textsuperscript{a}Department of Electrical and Electronic Engineering, Bangladesh University of Engineering and Technology, Dhaka, 1000, Bangladesh , \textsuperscript{b} Nanotechnology Research Laboratory, Department of Physics, Bangladesh University of Engineering and Technology, Dhaka-1000, Bangladesh.\\ \\ DOI: \href{https://doi.org/10.1016/j.jmrt.2021.06.026}{10.1016/j.jmrt.2021.06.026}}


\begin{abstract}
We have synthesized Gd\textsubscript{2}FeCrO\textsubscript{6} (GFCO) double perovskite which crystallized in monoclinic structure with P2$_1$/n space group. The UV-visible and photoluminescence spectroscopic analyses confirmed its direct band gap semiconducting nature. Here, by employing experimentally obtained structural parameters in first-principles calculation, we have reported the spin-polarized electronic band structure, charge carrier effective masses, density of states, electronic charge density distribution and optical absorption property of this newly synthesized GFCO double perovskite. Moreover, the effects of on-site d-d Coulomb interaction energy (U\textsubscript{eff}) on the electronic and optical properties were investigated by applying a range of Hubbard U\textsubscript{eff} parameter from 0 to 6 eV to the Fe-3d and Cr-3d orbitals within the generalized gradient approximation (GGA) and GGA+U methods. Notably, when we applied U\textsubscript{eff} in the range of 1 to 5 eV, both the up-spin and down-spin band structures were observed to be direct. The charge carrier effective masses were also found to enhance gradually from U\textsubscript{eff} = 1 eV to 5 eV, however, these values were anomalous for U\textsubscript{eff} = 0 and 6 eV. These results suggest that U\textsubscript{eff} should be limited within the range of 1 to 5 eV to calculate the structural, electronic and optical properties of GFCO double perovskite. Finally we observed that considering U\textsubscript{eff} = 3 eV, the theoretically calculated optical band gap $\sim$1.99 eV matched well with the experimentally obtained value $\sim$2.0 eV. The outcomes of our finding imply that the U\textsubscript{eff} value of 3 eV most accurately localized the Fe-3d and Cr-3d orbitals of GFCO keeping the effect of self-interaction error from the other orbitals almost negligible. Therefore, we may recommend U\textsubscript{eff} = 3 eV for first-principles calculation of the electronic and optical properties of GFCO double perovskite that might have potential in photocatalytic and related solar energy applications.
\end{abstract}


\maketitle

\section{Introduction} \label{sec:intro}
    Over the past few decades, the double perovskite oxides A$_{2}$BB$^{\prime}$O$_{6}$ (A = rare earth metal ions, B = transition metal ions) have gained immense research interest owing to their rich multifunctional characteristics and potential applications in the next generation spintronic devices \cite{Vasala2015,Gray2010,Gaikwad2019,Das2008,Ceron2019,Yin2019}. Recently,
several double perovskites have been reported to possess promising applicability in the field of photocatalysis, photovoltaic devices and photo(electro)chemical energy storage systems \cite{Lin2021,Mohassel2020,Kangsabanik2020,Yin2019}. Especially, A$_{2}$FeCrO$_{6}$ double perovskites (A = Pr, Bi etc.) having two 3d transition elements at B and B$^{\prime}$ sites have demonstrated fascinating optoelectronic properties such as favorable band gap energy, strong absorbance in the visible regime of the solar energy etc \cite{Wu2020,Gaikwad2019,Nechache2015}. Therefore, it is intriguing to investigate other members of A$_{2}$FeCrO$_{6}$ double perovskite family for enormous applications in electronics, photochemistry and optical technologies.

Nevertheless, it is difficult to synthesize perfectly ordered structure of A$_{2}$FeCrO$_{6}$ double perovskites because of the complementary ionic radii of Fe and Cr ions \cite{Gray2010,Nair2014}. Hence, the double perovskites of this family is yet less explored as compared to analogous double perovskite materials. Recently, we have successfully synthesized nanoparticles of double perovskite Gd$_{2}$FeCrO$_{6}$ (GFCO) for the first time by optimizing the synthesis condition of a citrate-based sol-gel technique and extensively investigated their crystallographic and chemical structure as well as  magnetic and optical behaviors \cite{Bhuyan2021}. Interestingly, the favorable surface morphology, optimal
direct band gap of $\sim$2.0 eV and the band edge positions of synthesized GFCO double perovskite have revealed its promising potential for visible light driven photocatalysis and related applications. Unfortunately, the complexity of experimental conditions and to some extent, the unavailability of the required experimental set-up posed difficulty to understand their optical and electronic characteristics at atomic level. Such impediment can be overcome by performing density functional theory (DFT) based first-principles calculation systematically \cite{Zhang2016}.

However, it should be noted that the standard DFT methods, for instance local density approximation (LDA) and generalized gradient approximation (GGA) of the exchange-correlation functional have some limitations in analyzing correctly the electronic properties of strongly correlated systems like GFCO \cite{Shenton2017}. To be specific, the LDA and GGA methods have demonstrated systemic failures to explain the on-site Coulomb interactions of highly localized electrons because of the erroneous electron self-interaction \cite{Himmetoglu2014}. One of the notable deficiencies of standard approximations is the underestimation of the optical band gap in semiconducting double perovskites which might have potentiality in photocatalytic and optoelectronic applications \cite{Terakura1984,Brown2020}. These limitations of standard DFT can be reasonably corrected by GGA+U method in which on-site Hubbard-like correction is applied to the effective potential \cite{Lu2014,Wang2006,Zhang2010}. Notably, two free parameters, U and J are required to effectively tune the on-site Coulomb and exchange interactions, respectively \cite{Shenton2017}. In an approach proposed by Dudarev \textit{et. al.} \cite{Dudarev1998}, these two parameters can be combined into a single Hubbard U\textsubscript{eff} correction parameter where U\textsubscript{eff} = U-J \cite{Wang2006}.

Typically, within GGA+U calculation, the U\textsubscript{eff} value is selected such that the calculated band gap matches with the experimentally obtained one \cite{Shenton2017}. However, a number of recent investigations \cite{Shenton2017, Perdew1986} have demonstrated that choosing U\textsubscript{eff} parameter only to match the band gap, may introduce spurious effects for strongly correlated materials. For instance, Shenton \textit{et al.} \cite{Shenton2017} reported that a U\textsubscript{eff} value of 5 eV or larger is required to match the experimental band gap of BiFeO\textsubscript{3}. However, the ordering of the Fe d orbitals at the conduction band minimum of BiFeO\textsubscript{3} inverts for U\textsubscript{eff} $>$ 4 eV. Hence, careful consideration is required to apply the most accurate Hubbard U\textsubscript{eff} parameter in GGA so that the theoretical band gap value closely matches with the experimental one without introducing any significant error in the character of electronic band edges. To the best of our knowledge, the influences of Hubbard U\textsubscript{eff} parameter on the optical and electronic properties of double perovskites possessing two 3d transition elements like GFCO have not been extensively investigated yet.

Therefore, in the current work, we have extensively investigated the effect of Hubbard U\textsubscript{eff} parameter on the crystallographic parameters, spin-polarized electronic band structure and optical properties of our recently synthesized GFCO nanoparticles by performing first-principles calculation via both GGA and GGA+U methods. To ensure reliability, we have employed our experimentally obtained structural parameters for theoretical analysis. We observed that the variation in U\textsubscript{eff} had insignificant effect on the structural parameters of GFCO. However, the character and curvature of electronic band edges could not be determined accurately without applying U\textsubscript{eff} i.e. for U\textsubscript{eff} = 0 eV and also, for U\textsubscript{eff} $>$ 5 eV in GGA+U calculation. Finally, considering the theoretically calculated optical band gap, we conjectured that a U\textsubscript{eff} value of 3 eV is reasonable to employ to the Fe-3d and Cr-3d orbitals of GFCO double perovskite within this calculation.

\section{Experimental and computational details}

Double perovskite GFCO nanoparticles were synthesized by adopting a standard citrate-based sol-gel technique, as discussed
in details in our previous work \cite{Bhuyan2021}. The crystallographic phase, lattice parameters, bond angles and bond lengths of as-synthesized GFCO were determined by Rietveld refinement of the powder XRD data using FullProf computer program package \cite{rodriguez1990fullprof}. An ultraviolet-visible (UV-visible) spectrophotometer (UV-2600, Shimadzu) was used to obtain absorbance spectrum of the as-synthesized GFCO for wavelengths ranging from 200 to 800 nm. Steady-state photoluminescence (PL) spectroscopy was conducted at room temperature by Spectro Fluorophotometer (RF-6000, Shimadzu). In the present investigation, based on our experimental findings, spin-polarized optical properties and accurate electronic band structure of as-prepared GFCO nanoparticles were determined theoretically by DFT based first-principles calculation.

\begin{figure}
\centering
\includegraphics[width=75mm, scale=0.8]{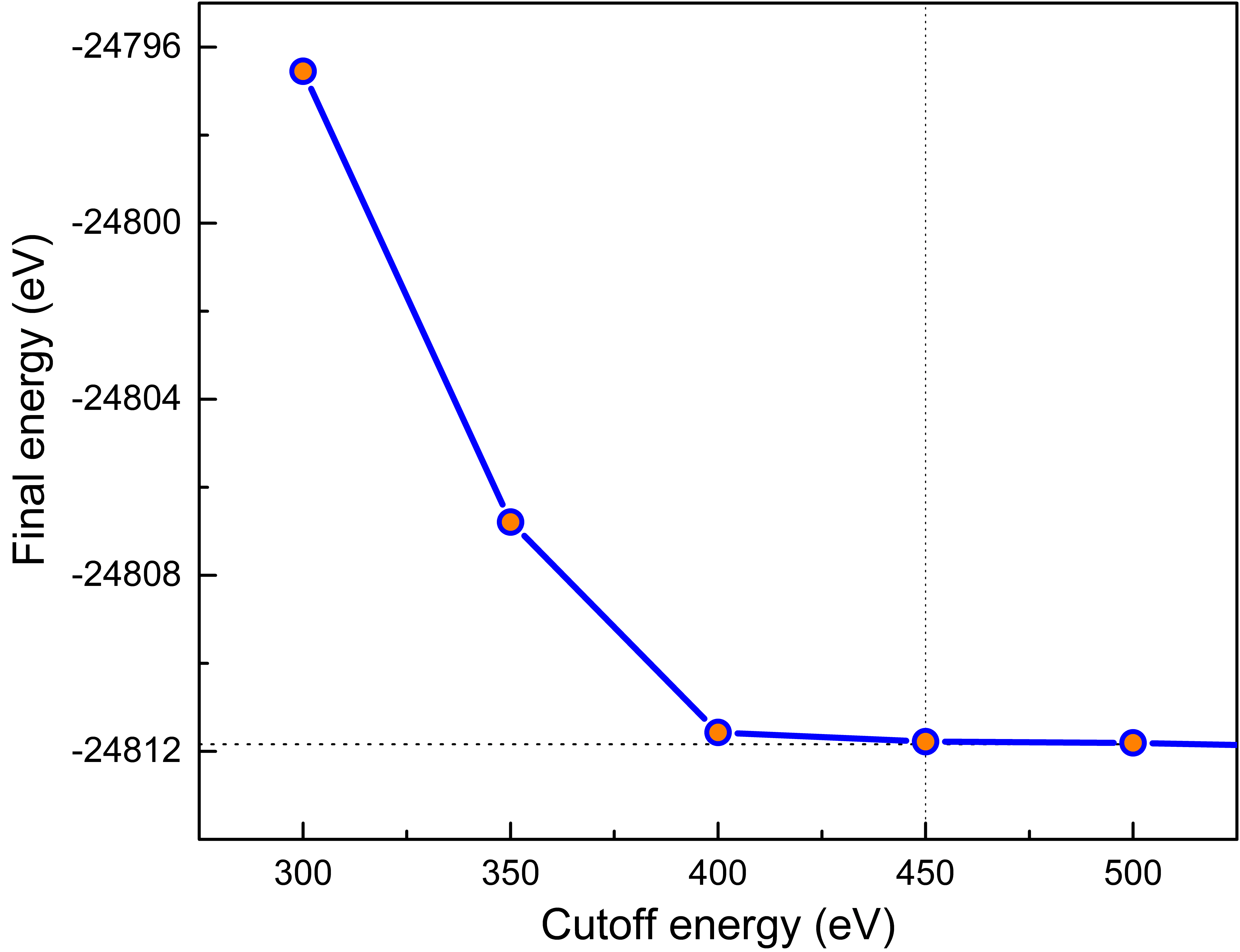}
\caption{\label{fig:epsart} Plane-wave cutoff energy convergence for structural optimization.}
\end{figure}

The theoretical calculations were carried out using both generalized gradient approximation (GGA) and GGA+U methods within the plane wave pseudopotential (PWPP) framework as implemented in the Cambridge Serial Total Energy Package (CASTEP) \cite{Segall2002, Rozilah2017}. The crystallographic structural parameters obtained from the Rietveld refined powder XRD spectrum of GFCO \cite{Bhuyan2021} were employed for DFT calculation. Prior to calculation, the geometry was optimized via Brodyden-Fletcher-Goldfarb-Shanno (BFGS) scheme applying energy of 10$^{-5}$ eV/atom, maximum force of 0.05 eV/Å and maximum stress of 0.1 GPa \cite{Fischer1992}. The Gd-4f$^{8}$5s$^{2}$5p$^{6}$6s$^{2}$, Fe-3d$^{6}$4s$^{2}$, Cr-3s$^{2}$3p$^{6}$3d$^{5}$4s$^{1}$ and O-2s$^{2}$2p$^{4}$ electrons were treated as valence electrons. The plane-wave cutoff energy convergence result for structural optimization is demonstrated in Fig. 1 where the dashed line represents the default energy cutoff. As can be observed, 450 eV energy cutoff was found sufficient to achieve the converged ground-state energy of GFCO. Hence, the plane-wave basis set was employed with the optimized energy cutoff of 450 eV. Moreover, Brillouin-zone integration were carried out with a 5$\times$5$\times$3 Monkhorst-Pack k-point mesh \cite{Monkhorst1976}. Spin polarized mode was endorsed during self-consistent field (SCF) calculations and a SCF tolerance of 2$\times$10$^{-6}$ eV per atoms was used.

Notably, to describe the exchange-correlation energy, at first we have used the GGA (U\textsubscript{eff} = 0 eV) method based on the Perdew-Burke-Ernzerhof functional (PBE) within on-the-fly generated  ultrasoft pseudopotentials (USP). Then a spin-polarized calculation was carried out to confirm the exact ground state of GFCO double perovskite \cite{Zhang2010}. Further, the GGA+U calculation was performed with different values of U\textsubscript{eff} to investigate the effect of Hubbard U\textsubscript{eff} parameter on the structural, electronic and optical properties of the GFCO ground state \cite{Dudarev1998}. To be specific, U\textsubscript{eff} was varied from 1 to 6 eV for Fe-3d and Cr-3d orbitals whereas for Gd-4f orbital, U\textsubscript{eff} was kept fixed at 6 eV in accordance with previous investigations \cite{Rozilah2017,Chakraborty2008}.
 
 The optical absorption coefficient was determined using the equation \( ~ \alpha =\sqrt[]{2} \omega \sqrt[]{~\sqrt[]{ \varepsilon _{1}^{2} \left(  \omega  \right) + \varepsilon _{2}^{2} \left(  \omega  \right)} - \varepsilon _{1} \left(  \omega  \right) } \), where  \(  \varepsilon _{1} \left(  \omega  \right)  \)  and  \(  \varepsilon _{2} \left(  \omega  \right)  \)  denote frequency dependent real and imaginary parts of dielectric function,  \(  \omega  \)  represents the photon frequency \cite{Saha2000}.   $\varepsilon _{1} \left(  \omega  \right)$    was calculated from the   $\varepsilon _{2} \left(  \omega  \right)$ by the Kramers-Kronig relationship \cite{Zhang2008}.

\section{Results and discussion}
\subsection{Crystal structure}

\begin{table*}
\centering
\caption{\label{tab:table3}Lattice parameters, monoclinic angle and unit cell volume of Gd$_2$FeCrO$_6$ for different values of U\textsubscript{eff} obtained via first-principles calculation along with the corresponding experimental values.}
\resizebox{\textwidth}{!}{
\begin{tabular}{lllllllll}
\hline

          & Experimental  value  & U\textsubscript{eff}= 0 eV & U\textsubscript{eff} = 1 eV & U\textsubscript{eff} = 2 eV & U\textsubscript{eff} = 3 eV & U\textsubscript{eff} = 4 eV & U\textsubscript{eff} = 5 eV & U\textsubscript{eff} = 6 eV \\ \hline
a ($\AA$)     & 5.359   & 5.397  & 5.442   & 5.448   & 5.457   & 5.466   & 5.470  & 5.502                 \\
b ($\AA$)     & 5.590 & 5.601  & 5.676   & 5.682   & 5.689   & 5.698   & 5.701  & 5.750                   \\
c ($\AA$)     & 7.675   & 7.688  & 7.781   & 7.798   & 7.808   & 7.820   & 7.821  & 7.886                 \\
$\beta$ ($^{\circ}$)    & 89.958    & 89.691 & 90.004  & 90.003  & 89.998  & 89.991  & 89.990 & 89.989                \\
Volume($\AA$$^{3}$)  & 229.92 & 232.43 & 240.35  & 241.42  & 242.46  & 243.73  & 243.85 & 249.54               \\
\hline

\end{tabular}}
\end{table*}

In our previous investigation \cite{Bhuyan2021}, we have extensively investigated the crystallographic structure of our as-synthesized GFCO nanoparticles at room temperature by performing Rietveld refinement analysis of their powder XRD pattern. Notably, it was observed that GFCO crystallizes in monoclinic structure with P2$_{1}$/n space group. The lattice parameters of GFCO unit cell were found to be $a$ = 5.359(1) $\AA$, $b$ = 5.590(2) $\AA$, $c$ = 7.675(3) $\AA$, monoclinic angle $\beta$ = 89.958(1)$^{\circ}$ with cell volume 229.920 $\AA^{3}$. In the present work, we have performed first-principles calculation to determine the structural parameters of GFCO by GGA (U\textsubscript{eff} = 0 eV) and  GGA+U (U\textsubscript{eff} = 1 to 6 eV) approaches. The calculated lattice constants $a$, $b$ and $c$, monoclinic angle ($\beta$) along with unit cell volume are tabulated in Table 1. For comparison, we have also included the experimentally obtained structural parameters in the table. Noticeably, the lattice constants and monoclinic angles obtained via first-principles calculation were found to be slightly larger than the experimental values. Nevertheless, this mismatch remained nominal i.e. within 3\% of the experimental results which is consistent with a number of previous investigations \cite{Shenton2017,Mann2016}. For instance, the lattice parameters of CO\textsubscript{2}-metal organic framework calculated with Hubbard U corrections were reported to remain within 3\% of experimental values \cite{Mann2016}. Moreover, it can be observed that the calculated lattice parameters and unit cell volume enhanced with the increment of U\textsubscript{eff} from 0 to 6 eV \cite{Shenton2017}. In contrast, the monoclinic angle obtained by the GGA+U method decreased with increasing U\textsubscript{eff} in the range of 1 to 6 eV. Such observations are in well agreement with previous investigations of related materials \cite{Shenton2017,Neaton2005}. 

\begin{figure}
\centering
\includegraphics[width=75mm]{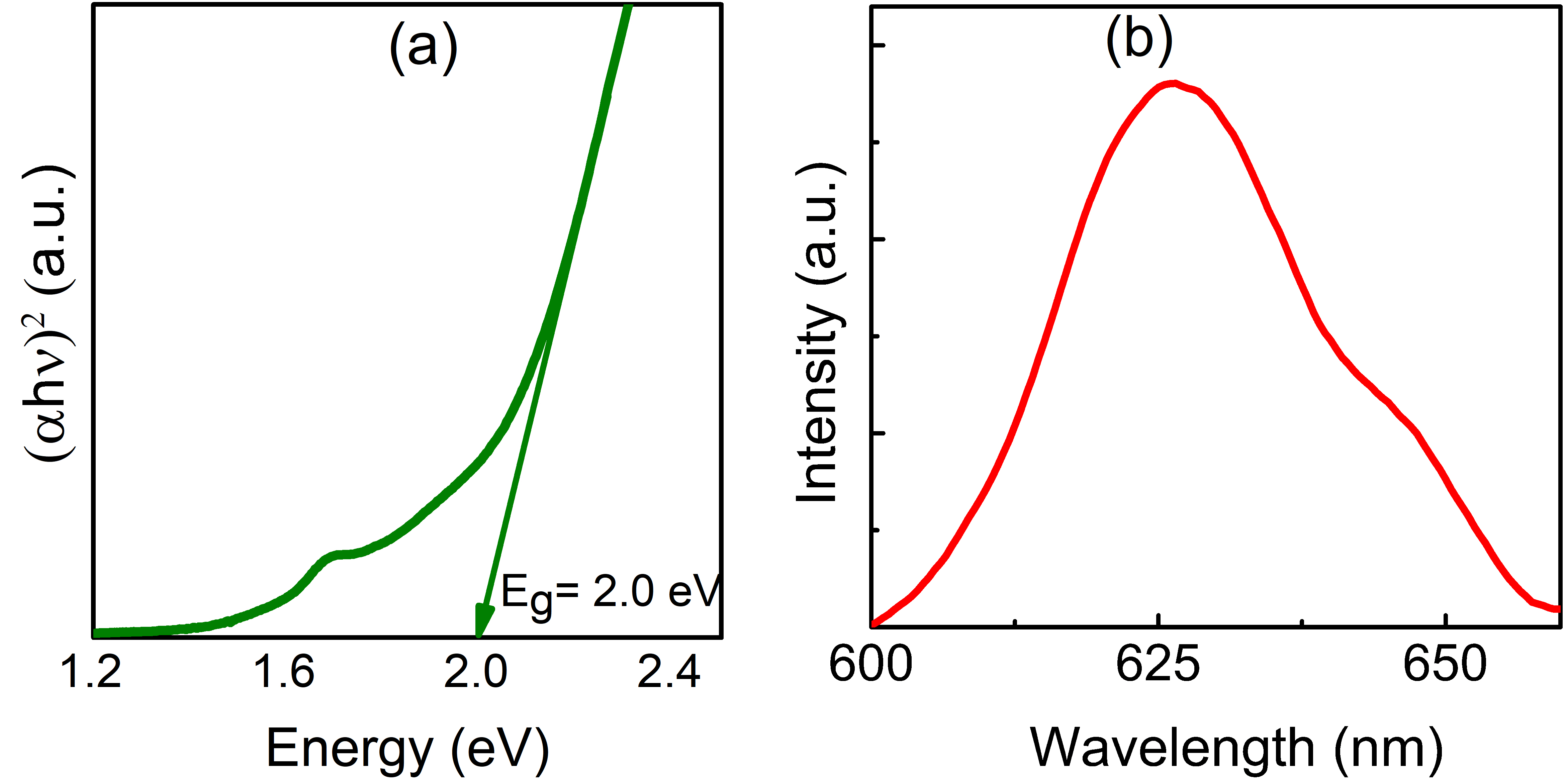}
\caption{\label{fig:epsart} Experimentally obtained (a) Tauc plot for direct optical band gap estimation and (b) steady-state photoluminescence spectrum of Gd$_2$FeCrO$_6$ nanoparticles \cite{Bhuyan2021}.}
\end{figure}

\subsection{Experimentally obtained optical properties}

The optical characteristics of as-synthesized GFCO nanoparticles were extensively investigated by obtaining their UV-visible absorbance spectrum and was reported previously \cite{Bhuyan2021}. The absorbance data was employed to calculate the optical band gap of synthesized GFCO double perovskite using Tauc relation \cite{Tauc1966}. The generated Tauc plot for estimating the direct optical band gap of GFCO nanoparticles is shown in Fig. 2(a). The abscissa intercept of the tangent to the linear region of the curve demonstrated that the optical band gap value is $\sim$2.0 eV. 

To further ensure the estimated direct optical band gap of GFCO perovskite, the steady-state PL spectrum of the synthesized material was recorded for an excitation wavelength of 230 nm \cite{Bhuyan2021}. The position of the PL peak in Fig. 2(b) confirmed that the band gap value of GFCO is $\sim$1.98 eV which closely matches with the direct band gap value obtained from the Tauc plot (Fig. 2(a)).  Therefore, both the UV-visible and PL spectroscopic analyses revealed that our as-synthesized nanostructured GFCO is a direct band-gap double perovskite with a band gap value of $\sim$2.0 eV. The experimentally observed optical properties of GFCO perovskite suggest the semiconducting nature of GFCO and most importantly, demonstrates its ability to absorb light of the visible spectrum of the solar illumination. \\

 \begin{figure*}
 \centering
\includegraphics[width=150mm]{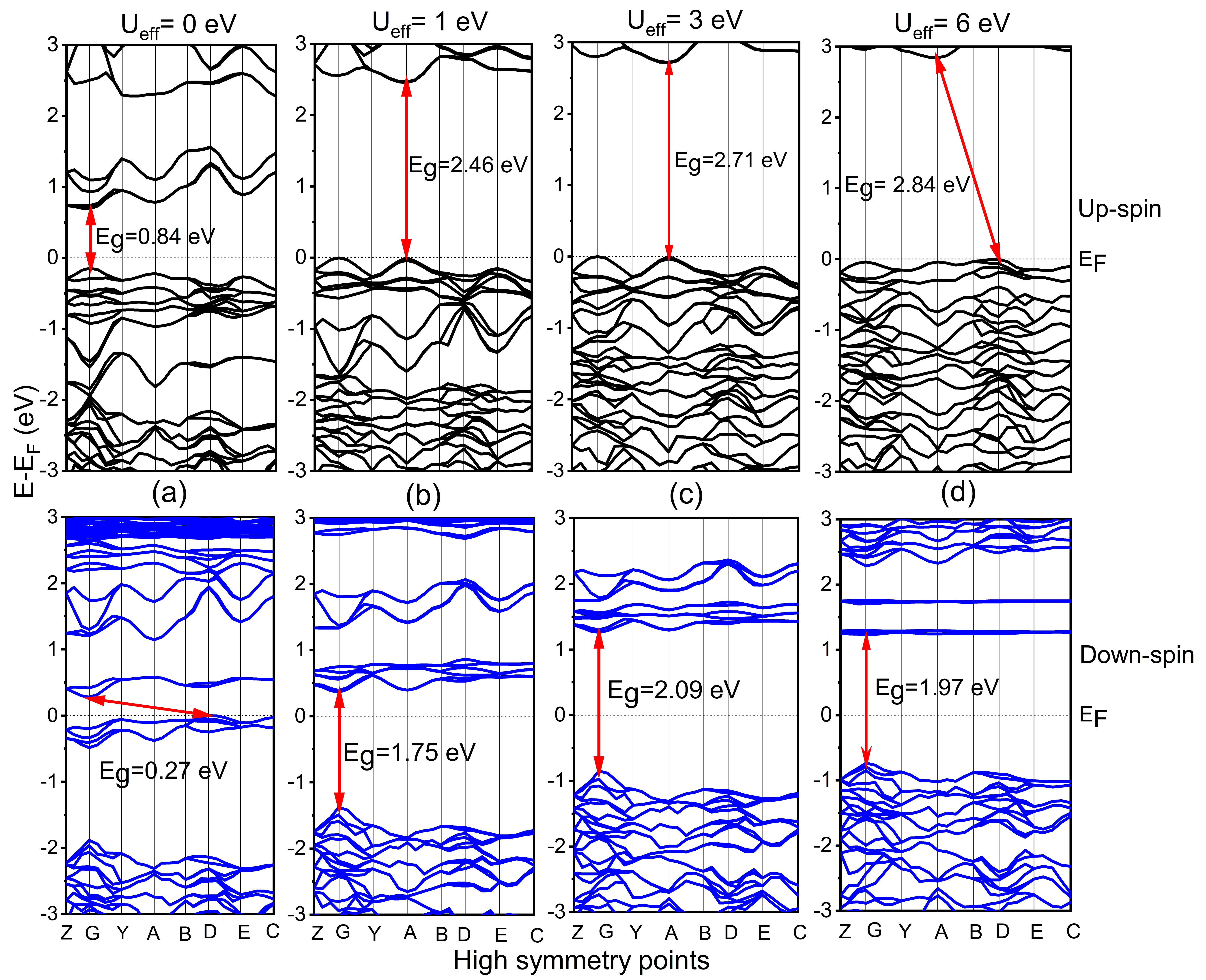}
\caption{\label{fig:epsart}  Electronic band structure of Gd\textsubscript{2}FeCrO\textsubscript{6}  for (a) U\textsubscript{eff} = 0 eV, (b) U\textsubscript{eff} = 1 eV, (c) U\textsubscript{eff} = 3 eV, and (d) U\textsubscript{eff} = 6 eV. Black and blue curves represent up-spin and down-spin orientations, respectively. The energy
ranges from -3 to 3 eV and the zero is set to the Fermi energy E\textsubscript{F}.}
\end{figure*}

\subsection{Theoretical investigation of electronic properties}
\subsubsection{Electronic band structure}

 The theoretical investigation was initiated by calculating the spin-polarized electronic band structure of GFCO within the GGA method (U\textsubscript{eff} = 0 eV) as shown in Fig. 3(a). The dotted horizontal line between the valence and conduction bands represents the Fermi level \cite{{Hou2014}}. As can be observed, for up-spin orientation, we obtained a direct electronic band gap of 0.84 eV whereas for down-spin, the band gap is found to be indirect having a value of 0.27 eV. Notably, both of these values are much smaller than the direct optical band gap $\sim$2.0 eV of as-synthesized GFCO as confirmed by UV-visible and PL spectroscopic analyses. It is well known that electronic band gap of any material would be larger than its optical band gap \cite{Bredas2014}, hence we may infer that the band structure formed for U\textsubscript{eff} = 0 eV is incorrect. Therefore, we have determined the electronic band structure of GFCO via GGA+U method with U\textsubscript{eff} = 1 to 6 eV. The electronic band structures obtained for U\textsubscript{eff} = 1, 3, 6 eV are presented in Fig. 3(b), (c) and (d), respectively and the band structures calculated for U\textsubscript{eff} = 2, 4, 5 eV are provided in  Fig. S1 of Electronic Supplementary Information (ESI). As can be seen in these two figures, the gap in the up-spin band enlarges with increasing U\textsubscript{eff} which can be attributed to the enhanced localization of the Fe-3d and Cr-3d orbitals due to increased U\textsubscript{eff} \cite{Shenton2017}. It is worth noting that for U\textsubscript{eff} = 1 to 5 eV, both the valence band maximum (VBM) and conduction band minimum (CBM) were within the A symmetry point indicating direct band structure. 

However, at U\textsubscript{eff} = 6 eV, we obtained an indirect up-spin band structure which is inconsistent with all the previous cases. It should be noted that so far both experimental and theoretical calculations provided strong evidence in support of direct band structure of GFCO. Therefore, the indirect band structure obtained for U\textsubscript{eff} = 6 eV calls into question the applicability of employing large Hubbard parameter (i.e. U\textsubscript{eff} $>$ 5 eV) in first-principles calculations of GFCO double perovskite \cite{Shenton2017}. Moreover, it is interesting to note that the theoretically calculated up-spin band gap values were within the range of 2.46 to 2.84 eV which ensures the semiconducting nature of GFCO double perovskite as was also evident from the UV-visible and PL spectroscopic analyses (Fig. 2). 

In the case of down-spin orientation, for all values of U\textsubscript{eff} (1 to 6 eV), both the CBM and VBM were obtained within the G symmetry point indicating again direct band structure. Interestingly,  the band gap value increased monotonically from U\textsubscript{eff} = 1 to 5 eV but an anomalous decrease can be observed at U\textsubscript{eff} = 6 eV (Fig. 3(d)). Therefore, it might be conjectured that the optimized value of U\textsubscript{eff} for first-principles calculation of GFCO would be less than 6 eV \cite{Shenton2017}.

\begin{figure}
\centering
\includegraphics[width=75mm]{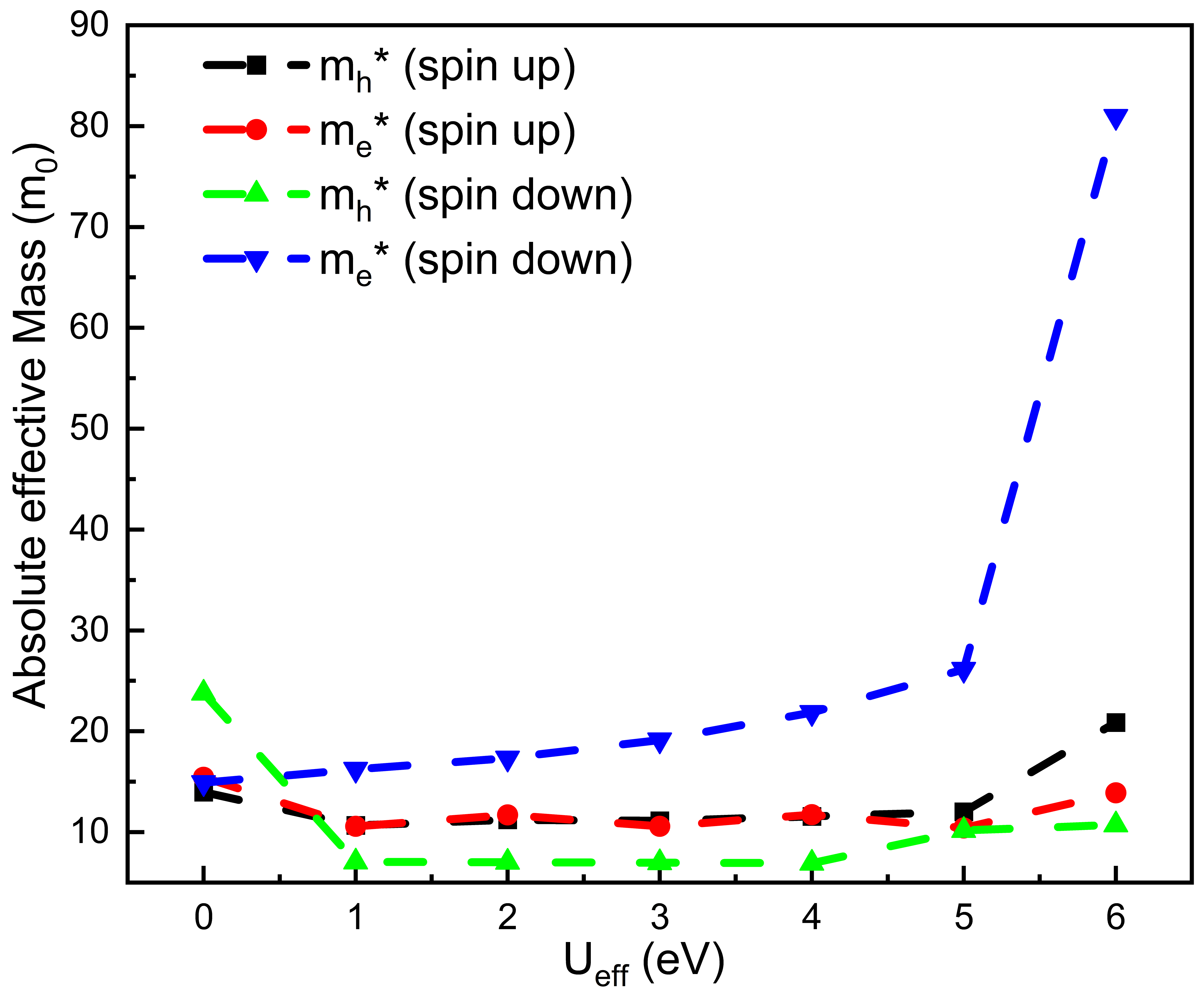}
\caption{\label{fig:epsart}  Variation in absolute charge carrier effective mass as a function of U\textsubscript{eff}. m$_{e}$$^{*}$ and m$_{h}$$^{*}$ are the electron and hole effective masses, respectively in units of the electron rest mass, m$_{o}$ } 
\end{figure}

For understanding the carrier transport in the material, we have quantified the curvature at band extrema by calculating the technologically important charge carrier effective masses using following expression \cite{Reunchan2016}.
\begin{equation}
m^{*}=\hbar^{2} \left ( \frac{\mathrm{d}^{2}E}{\mathrm{d} k^{2}} \right )^{-1}
\end{equation}
Here, E is the band-edge energy as a function of wave-vector k. The electron effective mass (m$_{e}$$^{*}$) was calculated by parabolic fitting of the E-k curve within the small
region of wave-vector near the CBM \cite{Reunchan2016,Das2019}. The hole effective mass (m$_{h}$$^{*}$) was estimated by analyzing the region near the VBM using similar approach \cite{Reunchan2016}. The variations of m$_{e}$$^{*}$ and m$_{h}$$^{*}$ of GFCO double perovskite as a function of U\textsubscript{eff} are illustrated in Fig. 4 both for up-spin and down-spin orientations. For up-spin band structure, we can observe a reduction in m$_{e}$$^{*}$ from 15.5m\textsubscript{o} to 10.6m\textsubscript{o} for U\textsubscript{eff} = 0 eV to U\textsubscript{eff} = 1 eV suggesting an increase in curvature at the CBM. Between U\textsubscript{eff} of 1 to 5 eV, the values of m$_{e}$$^{*}$ changed nominally ($\sim$1.3m\textsubscript{o}). Further, for U\textsubscript{eff} = 6 eV, an enhancement of 3.5m\textsubscript{o} can be noticed which corresponds to the reduction in curvature at CBM. Clearly, in Fig. 4, we can observe a similar dependence of m$_{h}$$^{*}$ on U\textsubscript{eff} both for up and down-spin orientations. Moreover, for down-spin band structure, a notable increase can be noticed in m$_{e}$$^{*}$ from 26.1m\textsubscript{o} to 81m\textsubscript{o} for U\textsubscript{eff} = 5 eV to U\textsubscript{eff} = 6 eV. However, the variation was comparatively smaller ($\sim$ 11.2m\textsubscript{o}) in the range of U\textsubscript{eff} = 0 to 5 eV. It is worth mentioning that while varying the values of U\textsubscript{eff} for electronic band structure calculation, we had kept the structural parameters fixed at experimentally obtained values. Hence, we can confirm that the variations in m$_{e}$$^{*}$ and m$_{h}$$^{*}$ with U\textsubscript{eff} are solely due to the electronic effects. Such variation trend again justifies that it would be worthwhile to consider U\textsubscript{eff} within the range 1 to 5 eV for GFCO double perovskite.

\subsubsection{Density of states}

\begin{figure*}
\centering
\includegraphics[width=150mm]{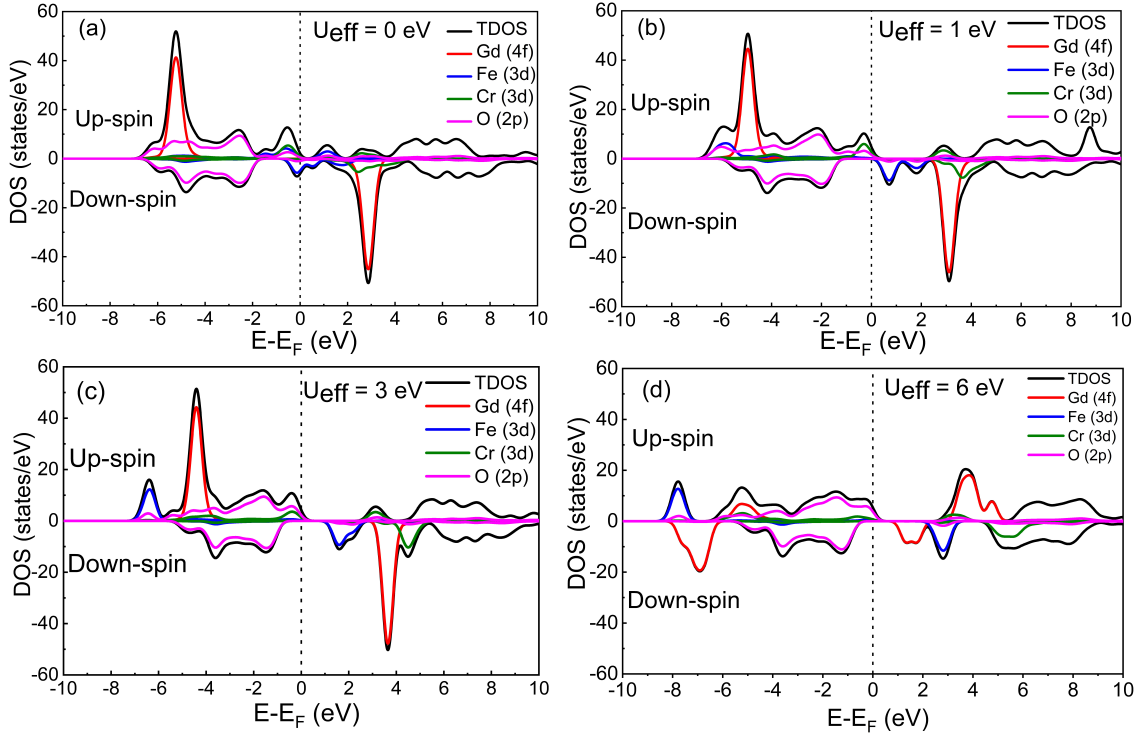}
\caption{\label{fig:epsart} Calculated total density of states (TDOSs) and partial density of states (PDOSs) of Gd-4f, Fe-3d, Cr-3d and O-2p orbitals for both up-spin and down-spin channels. The panels (a)–(d) show the DOSs for U\textsubscript{eff} = 0, 1, 3, 6 eV, respectively. The zero is set to the Fermi energy.}
\end{figure*}

In order to resolve the contribution of each individual orbital to the electronic bands of GFCO double perovskite, we have calculated the total density of states (TDOS) and partial density of states (PDOS) for Gd-4f, Fe-3d, Cr-3d and O-2p orbitals via GGA (U\textsubscript{eff} = 0 eV) and GGA+U methods (U\textsubscript{eff} = 1 to 6 eV). The DOSs obtained for U\textsubscript{eff} = 0, 1, 3, 6 eV are presented in Fig. 5(a), (b), (c) and (d), respectively and the DOSs calculated for U\textsubscript{eff} = 2, 4, 5 eV are provided in  Fig. S2 of ESI.

Beginning with the character of the conduction band (E-E\textsubscript{F}$>$ 0 eV) obtained for up-spin orientation, we can make the following observations. At U\textsubscript{eff} = 0 eV, Fig. 5(a), the conduction band lying at around 1 eV was made up of a hybridization of Fe-3d and O-2p orbitals. This band disappeared after considering the effect of Hubbard U\textsubscript{eff} which corresponds to the decrement in m$_{e}$$^{*}$ at U\textsubscript{eff} = 1 eV (Fig. 5(b)). Notably, the conduction band (at around 2.5 eV) obtained for U\textsubscript{eff} = 1 eV had primarily the characteristics of Cr-3d with a minor contribution from O-2p. Further, as shown in Fig. 5(c) and ESI Fig. S2, with the increase of U\textsubscript{eff} up to 5 eV, this conduction band shifted to higher energy resulting in the enlargement of band gap. Also, the contribution of Cr-3d to the up-spin conduction band enhanced with increasing U\textsubscript{eff}. Interestingly, at U\textsubscript{eff} = 6 eV (Fig. 5(d)), we observed that the computed PDOS for Gd-4f orbital radically altered and the conduction band arose mainly due to the hybridization of Gd-4f with a minor contribution from Cr-3d orbital. Such anomaly is another indication for the limitation of U\textsubscript{eff} $>$ 5 eV to explain the electronic band structure of GFCO double perovskite accurately.

Now, if we analyze the valence bands (E-E\textsubscript{F}$<$ 0 eV) of up-spin electronic band structure shown in Fig. 5(a), it can be observed that the valence band at U\textsubscript{eff} = 0 eV was composed of Cr-3d, O-2p as well as Fe-3d orbitals. However, when we employed U\textsubscript{eff}, the contribution of Fe-3d got diminished. To be specific, the up-spin valence band for U\textsubscript{eff} = 1 eV was made up of the hybridization of Cr-3d and O-2p states (Fig. 5(b)). From Fig. 5(c), (d) and ESI Fig. S2, it can be noticed that with increasing U\textsubscript{eff}, the contribution of Cr-3d state gradually decreased leaving only O-2p to dominate the valence band at U\textsubscript{eff} = 6 eV which can be associated with the enhancement of m$_{h}$$^{*}$.

Further Fig. 5(a) demonstrates that at U\textsubscript{eff} = 0 eV, the conduction band (E-E\textsubscript{F}$>$ 0 eV) of down-spin structure was dominated mostly by Fe-3d orbital with a negligible contribution from O-2p. As can be noticed in Fig. 5(b), (c) and ESI Fig. S2, the characteristics of the conduction band did not change much due to the effect of Hubbard U\textsubscript{eff} up to 5 eV. Only a gradual shifting of conduction band to a higher energy can be noticed with increasing U\textsubscript{eff} as expected \cite{Shenton2017}. However, when we employed a U\textsubscript{eff} of 6 eV (Fig. 5(d)), a new flat conduction band appeared at around 1.25 eV owing to the sole contribution of  Gd-4f state. Furthermore, in the case of the down-spin valence band (E-E\textsubscript{F}$<$ 0 eV), a significant influence of U\textsubscript{eff} can be observed. Without considering the Coulomb repulsion effect (i.e. for U\textsubscript{eff} = 0 eV in Fig. 5(a)), we obtained the valence band near the Fermi level which is attributed to the hybridization of Fe-3d and O-2p orbitals. After applying U\textsubscript{eff} = 1 eV (Fig. 5(b)), this band disappeared followed by the emergence of a new band at around -1.5 eV which shifted gradually to higher energy with increasing U\textsubscript{eff} of up to 6 eV (Fig. 5(b)-(d) and ESI Fig. S2). Notably, the formation of this band has the contribution from only O-2p orbital.

\subsubsection{Mulliken population analysis}

\begin{table*}[]
\caption{ Mulliken effective charges of individual atoms,
bond populations and bond lengths of Gd$_2$FeCrO$_6$ for different values of U\textsubscript{eff} obtained via Mulliken population analysis.}
\begin{tabular}{cccccccc}
\hline
     & U\textsubscript{eff} =   0 eV & U\textsubscript{eff} =1   eV & U\textsubscript{eff} = 2 eV & U\textsubscript{eff} = 3 eV & U\textsubscript{eff} = 4 eV & U\textsubscript{eff} = 5 eV & U\textsubscript{eff} = 6 eV \\ \hline
Atom & \multicolumn{6}{c}{Mulliken   effective charge (e)}     \\ \hline
Gd   & 1.52         & 1.49        & 1.49 & 1.49   & 1.49 & 1.50 & 1.50         \\
Fe   & 0.59         & 0.78        & 0.79 & 0.80   & 0.82    & 0.82  & 0.84         \\
Cr   & 0.53         & 0.58  & 0.59   & 0.61  & 0.62 & 0.63       & 0.65         \\
O    & -0.69        & -0.72       & -0.73   & -0.73 & -0.74 & -0.74     & -0.75        \\ \hline
Bond & \multicolumn{6}{c}{Bond   population}                   \\ \hline
Gd-O & 0.1450       & 0.1587     & 0.1600 &  0.1625  & 0.1637 & 0.1662  & 0.1700       \\
Fe-O & 0.3200       & 0.3100     & 0.3166 & 0.3166 & 0.3166 & 0.3166      & 0.3066       \\
Cr-O & 0.3533       & 0.3200   & 0.3200   & 0.3200 & 0.3200 & 0.3133      & 0.3100       \\
O-O  & -0.0350      & -0.0333  & -0.0333 &  -0.0333  & -0.0333 & -0.0333     & -0.0333      \\ \hline
Bond & \multicolumn{6}{c}{Bond   length (Å)}                   \\ \hline
Gd-O & 2.475        & 2.501   & 2.504 &  2.505 & 2.507 & 2.507    & 2.524        \\
Fe-O & 2.023        & 2.051       & 2.061 & 2.065  & 2.065 & 2.069     & 2.083        \\
Cr-O & 2.026        & 2.036      & 2.042 &   2.049 & 2.050 & 2.057      & 2.079      \\\hline 
\end{tabular}
\end{table*}

The effective atomic charge, bond population and bond length in a crystalline solid can be obtained from Mulliken population analysis which provides insight into the distribution of electrons among different parts of the atomic bonds, covalency of bonding as well as bond strength \cite{Mulliken1955,Segall1996}. Mulliken effective charge, Q($\alpha$) of a particular atom $\alpha$ can be calculated using the following expression \cite{Mulliken1955}.

\begin{equation}
    Q(\alpha )=\sum_{k}\omega _{k} \sum \sum_{\mu }^{on\;\alpha }\sum_{v}P_{\mu \nu }(k)S_{\mu\nu}(k)
\end{equation}
where P$_{\mu \nu }$ denotes an element of the density matrix and S$_{\mu \nu }$ refers to the overlap matrix.
And the bond population, P($\alpha\beta$) between two atoms $\alpha$ and $\beta$ can be expressed as \cite{Mulliken1955}-
\begin{equation}
    P(\alpha\beta  )=\sum_{k}\omega _{k} \sum \sum_{\mu }^{on\;\alpha }\sum_{v}^{on\;\beta }2P_{\mu \nu }(k)S_{\mu\nu}(k)
\end{equation}

Table II provides the calculated Mulliken effective charges of individual atoms, bond populations and bond lengths between different atoms in GFCO crystal structure. Noticeably, for all values of U\textsubscript{eff}, the Mulliken effective charges of the individual Gd, Fe, Cr and O atoms are found to be reasonably smaller than their formal ionic charges which are +3, +3, +3, and -2, respectively. Such difference between Mulliken effective and formal ionic charges is an indication of the existence of mixed ionic and covalent bonding in GFCO \cite{Segall1996}. It should be noted that small Mulliken effective charge of an atom is associated with its high level of covalency and vice versa \cite{Mulliken1955,Rozilah2017}. Therefore it can be inferred that GFCO double perovskite includes chemical bonding with prominent covalency. Further, in Table II, an enhancement can be observed in the effective charges of Fe, Cr and O atoms after employing U\textsubscript{eff} in the first-principles calculation (U\textsubscript{eff} =1 to 6 eV). This outcome indicates that the degree of bond covalency in GFCO reduced to an extent due to the effect of on-site Coulomb interaction. 

Table II also presents the bond populations and bond lengths of Gd–O, Fe–O, Cr–O and O-O bonds in GFCO double perovskite as obtained for U\textsubscript{eff} = 0 to 6 eV. It is noteworthy that a large positive value of bond population is associated with high degree of covalency whereas a small bond population indicates high degree of ionicity in the chemical bond \cite{Ching1999}. In the present investigation, the bond populations of the Gd–O, Fe–O and Cr–O were determined to be positive whereas the bond population of O-O was negative for all values of U\textsubscript{eff}. This result suggests that no bonds had formed between the O atoms in GFCO double perovskite \cite{Chakraborty2008,Yaakob2015}. Moreover, the calculated bond populations of Fe-O and Cr-O are found to be considerably larger than that of Gd-O. Such observation implies that the Fe-O and Cr-O bonds possess higher degree of covalency as compared to Gd-O bonds. It is also worth noticing that the bond populations of Fe-O and Cr-O calculated by GGA+U method (U\textsubscript{eff} =1 to 6 eV) are lower than the values obtained by GGA method (U\textsubscript{eff} = 0 eV). This result provides further evidence for the influence of Coulomb repulsion to reduce the covalency of Fe-O and Cr-O bonding in GFCO as was observed before by analyzing the calculated Mulliken effective charges. Furthermore, Table II demonstrates that the bond lengths of Fe-O and Cr-O are reasonably smaller than that of Gd-O which can be attributed to their larger bond population and consequently, higher degree of covalency as compared to Gd-O bonding.

\subsubsection{Electron charge density}

\begin{figure}[b]
\centering
\includegraphics[width=65mm, scale=0.8]{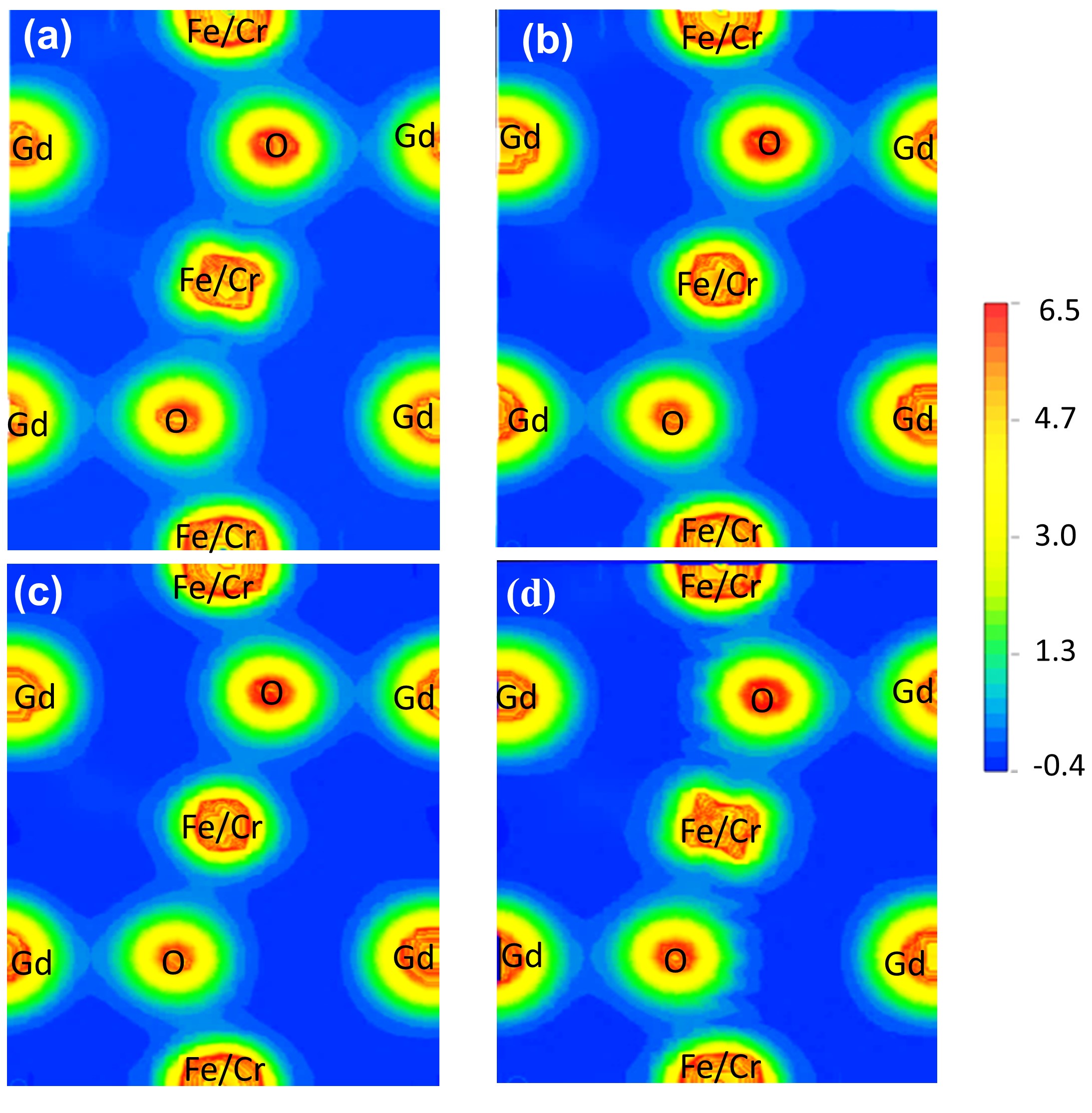}
\caption{\label{fig:epsart} Electronic charge density along z-axis of Gd$_2$FeCrO$_6$ for (a) U\textsubscript{eff} = 0 eV, (b) U\textsubscript{eff} = 1 eV, (c) U\textsubscript{eff} = 3 eV, and (d) U\textsubscript{eff} = 6 eV. }
\end{figure}

For further understanding of the chemical bonding nature, we have determined the electron charge density distribution of GFCO double perovskite along the z-axis by GGA (U\textsubscript{eff} = 0 eV) and GGA+U (U\textsubscript{eff} =  1, 3 and 6 eV) methods as illustrated in Fig. 6. It is worth noting that a typical covalent bond between two atoms involves overlapping of electron clouds from both of them and the electrons remain concentrated in the overlapping region \cite{Phillips1968}. In Fig. 6, for all values of U\textsubscript{eff}, a larger overlap of electron cloud can be observed between Fe/Cr and O atoms as compared to Gd and O atoms. Such observation implies that the covalent bonds between Fe/Cr and O in GFCO are considerably stronger than the bond between Gd and O as was also revealed by Mulliken population analysis. 

Moreover, it can be clearly seen that no electron clouds are concentrated in the area between one of the two Gd atoms and O which implies the formation of an ionic bond between these two atoms \cite{Craig1954,Rozilah2017}. Notably, the charge sharing between Fe/Cr-O can be attributed to the hybridization among Fe/Cr-3d and O-2p orbitals and the Gd-O bond formation is associated with the hybridization between Gd-4f and O-2p orbitals which was demonstrated by the DOS curves (Fig. 5).

Furthermore, if we meticulously observe Fig. 6(a) and (b), it can be noticed that the overlapping of electron clouds between Fe/Cr and O atoms got reasonably narrower after considering the effect of on-site Coulomb interaction (U\textsubscript{eff}) in first-principles calculation. Afterward, the electron charge density in the overlapped region between Fe/Cr and O atoms enhanced with increasing U\textsubscript{eff} (see Fig. 6(c) and (d)) which is consistent with the outcome of Mulliken population analysis.

\begin{figure*}
\centering
\includegraphics[width=150mm]{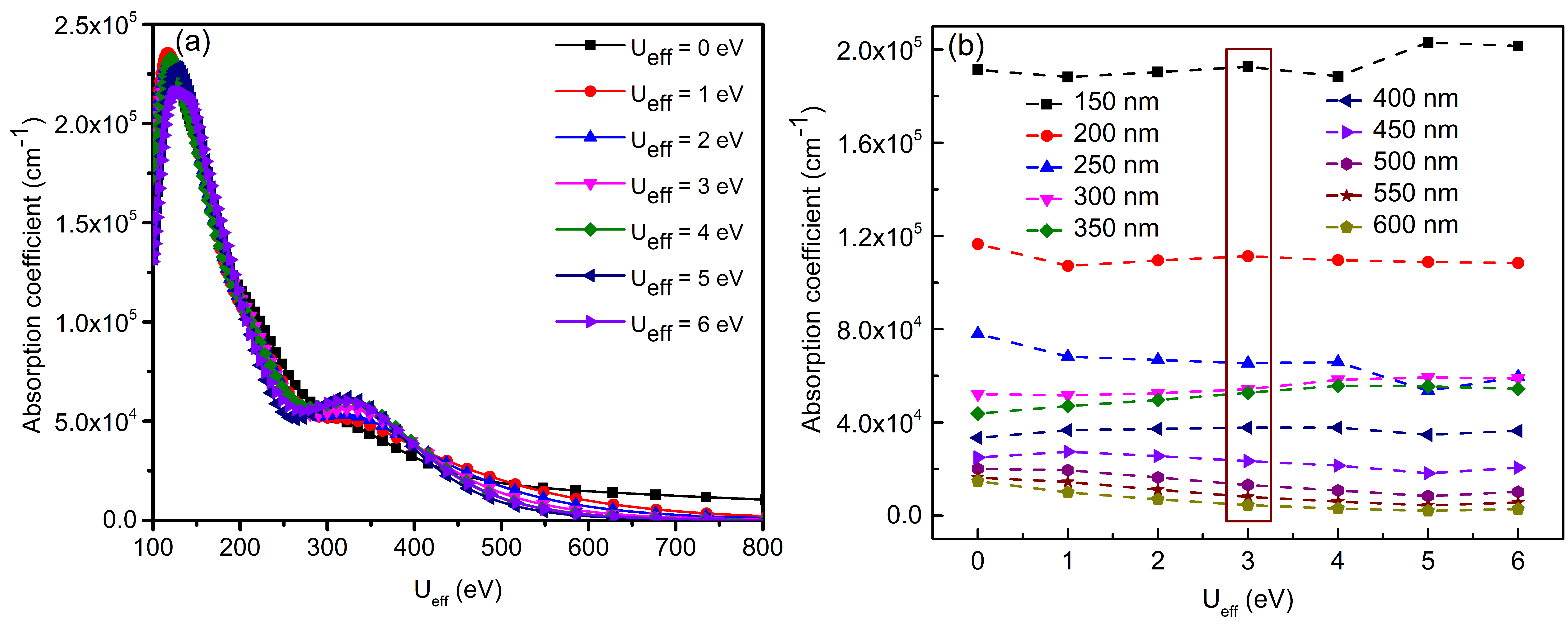}
\caption{\label{fig:epsart} (a) Variation of theoretically obtained absorption coefficient of GFCO perovskite as a function of wavelength for different U\textsubscript{eff}. (b) The absorption coefficient vs.  U\textsubscript{eff} for some fixed values of the wavelength.}
\end{figure*}

\subsection{Light absorption property}

The absorption coefficient provides valuable information about a material's light-harvesting ability. The optical absorption coefficient of GFCO has been evaluated by first-principles calculation via GGA (U\textsubscript{eff} = 0 eV) and GGA+U (U\textsubscript{eff} = 1 to 6 eV) approaches using a polarization technique which includes the electric field vector as an isotropic average in all directions \cite{Brik2014}. To gain additional distinguishable absorption peaks, a small smearing value of 0.5 eV was used. Fig. 7(a) illustrates the calculated absorption coefficients of GFCO double perovskite as a function of wavelength to demonstrate the effect of Hubbard U\textsubscript{eff} parameter on its light absorption property. In Fig. 7(a), for all values of U\textsubscript{eff}, two absorption peaks can be clearly observed in the UV region which indicates the strong UV light absorption capacity of GFCO. Noticeably, the stronger absorption coefficient peak lies at around 120 nm and it undergoes a slight red-shift with increasing U\textsubscript{eff}. On contrary, the weaker absorption peak at around 320 nm is slightly blue-shifted for higher values of U\textsubscript{eff}. For comparison, we have also provided the absorbance spectrum of GFCO obtained via UV-visible spectroscopy in ESI Fig. S3 \cite{Bhuyan2021}. It is worth noticing that the experimentally obtained spectrum has two additional bands in the visible region along with the two bands we obtained theoretically in the UV regime. This might be related to the fact that the DFT calculations were performed for 0 K whereas the experiment was conducted at room temperature. Such difference in conditions can be attributed to the discrepancy between the optical and theoretical absorbance spectra of GFCO \cite{honglin2014experimental,harun2020dft+}.

Fig. 7(b) shows the absorption coefficient vs. U\textsubscript{eff} curves of GFCO for some fixed values of wavelength ranging from 150 to 600 nm. We can observe that for the wavelength of 150 nm, the absorption coefficient attains its minimum and maximum values for U\textsubscript{eff} = 4 eV and U\textsubscript{eff} = 5 eV, respectively suggesting the underestimation and overestimation of the coefficient at these U\textsubscript{eff} values. Similarly,  it can be found that except at U\textsubscript{eff} = 3 eV, the absorption coefficient is either overestimated or underestimated at all other values of U\textsubscript{eff} for any of the specific wavelengths such as 200, 250, 300 nm etc. This intriguing observation is highlighted by the rectangle in Fig. 7(b).

\subsection{Comparison of experimental and theoretical optical band gaps}

Further, we have employed the calculated absorption coefficients to theoretically calculate the optical band gap values of GFCO double perovskite using Tauc relation \cite{Tauc1966}. Fig. 8 shows the variation in theoretically calculated direct band gap values as a function of U\textsubscript{eff}. Noticeably, a direct band gap value of 0.5 eV was obtained by GGA method (U\textsubscript{eff} = 0 eV) which is significantly smaller than the experimental value $\sim$2.0 eV. Further, with the increase of U\textsubscript{eff} up to 5 eV, an almost linear increase can be observed in the direct optical band gap values of GFCO double perovskite. However, an anomalous decrease can be noticed for a further increase of U\textsubscript{eff} to 6 eV. It is intriguing to note that the direct optical band gap ($\sim$1.99 eV) obtained for U\textsubscript{eff} = 3 eV matches well with the experimentally obtained one ($\sim$2.0 eV) which is marked by a circle in Fig. 8. Also, for U\textsubscript{eff} = 6 eV, we found the calculated band gap value ($\sim$2.08 eV) to be quite close to the experimental one. But as we had demonstrated before, the character and curvature of the conduction and valence bands inaccurately changed for U\textsubscript{eff} $>$ 5 eV. Hence, fitting U\textsubscript{eff} to the band gap alone may provide erroneous results for such double perovskite materials especially in the cases where the role of the band edge character is crucial.

Finally, it is fascinating to note that our predictions for all the properties of GFCO (i.e. structural, electronic and optical) considering U\textsubscript{eff} = 3 eV closely matched with the experimental results without any over or underestimation of band gap values. This implies that the U\textsubscript{eff} value of 3 eV most accurately localized the Fe-3d and Cr-3d orbitals of GFCO. Moreover, the almost accurate estimation of band gap values suggests that the effect of self-interaction error from the other orbitals of GFCO was almost negligible \cite{Brown2020, Perdew1986}. Therefore we recommend U\textsubscript{eff} = 3 eV for the GGA+U calculation of electronic and optical properties of GFCO double perovskite.

\begin{figure}
\centering
\includegraphics[width=75mm]{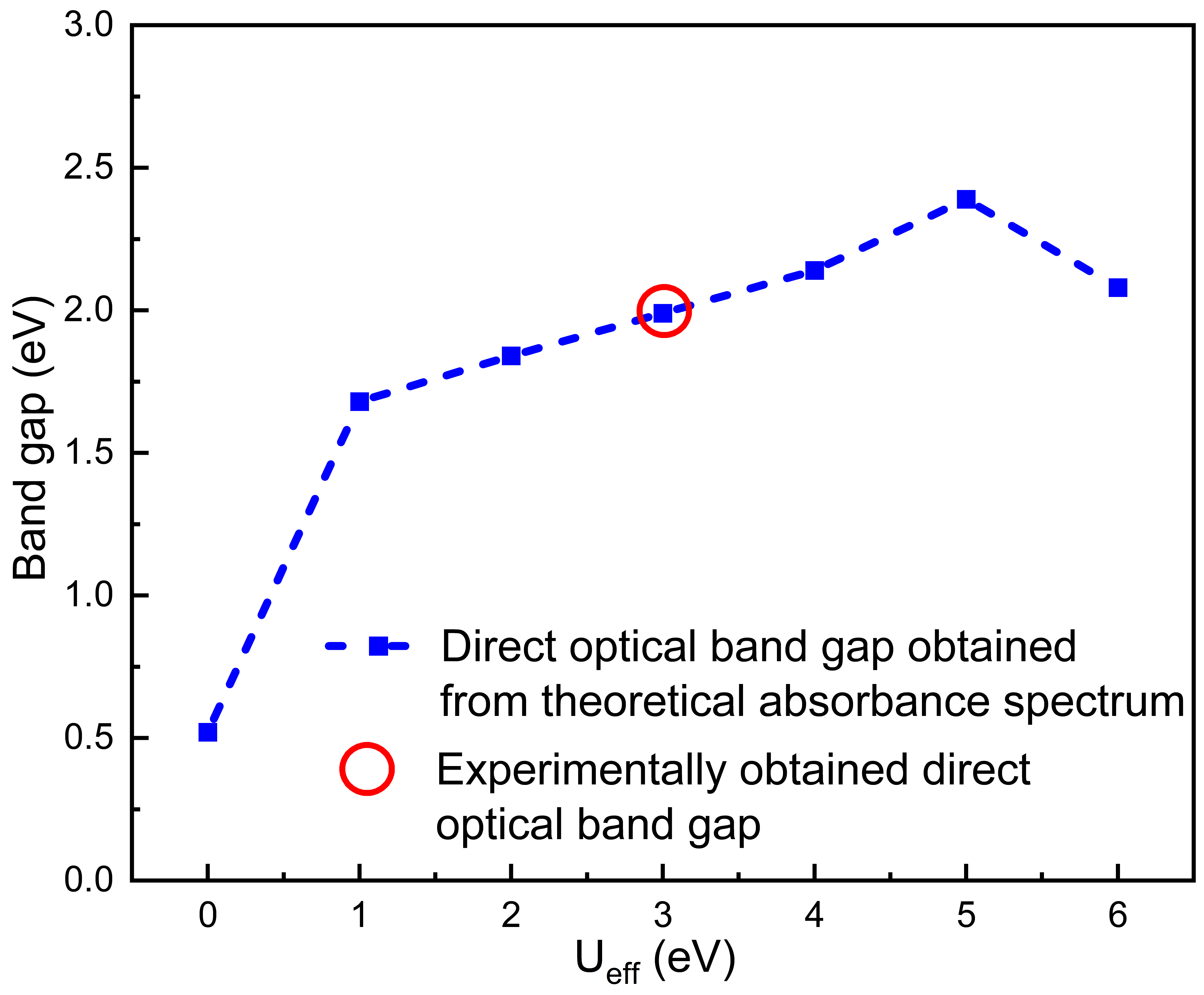}
\caption{\label{fig:epsart} Variation in theoretically calculated direct optical band gap as a function of U\textsubscript{eff}. The red circle represents the experimentally obtained optical band gap value.}
\end{figure}

\section{Conclusions}

We have employed the first-principles based GGA and GGA+U methods to calculate the spin-polarized electronic band structure, Mulliken bond population, electron charge density distribution and optical characteristics of our newly synthesized direct band-gap semiconductor Gd\textsubscript{2}FeCrO\textsubscript{6} (GFCO) double perovskite for a range of U\textsubscript{eff} between 0 and 6 eV, applied to the Fe-3d and Cr-3d orbitals. The structural parameters of the monoclinic GFCO crystal varied nominally with U\textsubscript{eff}. To the contrary, the variation of U\textsubscript{eff} demonstrated significant effect on the electronic band structure which indicates the importance of employing reasonable value of U\textsubscript{eff} to correct the over-delocalization of the Fe/Cr-3d states. For U\textsubscript{eff} $>$ 5 eV, the computed partial density of states for Gd-4f orbital radically altered which had significantly changed the band structure. In particular, we observed that the character and curvature of the conduction and valence bands largely varied for U\textsubscript{eff} $>$ 5 eV leading to enormous changes in calculated charge carrier effective masses. Notably, in the case of U\textsubscript{eff} = 3 eV, the theoretically calculated direct optical band gap $\sim$1.99 eV matched well with the experimental value $\sim$2.0 eV. These findings justify that it might be worthwhile to employ U\textsubscript{eff} = 3 eV to accurately calculate the structural, electronic and optical properties of GFCO double perovskite. The outcome of this investigation might be useful for a keen understating of the electronic and optical properties of this newly synthesized double perovskite and related material systems for photocatalytic applications via band gap engineering. This study also reveals the importance of conducting systematic analysis of the influence of on-site Coulomb interaction on band gaps as well as on the electronic structure as a whole for related other double perovskite materials for which experimental data are still not available.

\section*{Acknowledgements}
The computational facility provided by the Institute of Information and Communication Technology (IICT), Bangladesh University of Engineering and Technology (BUET) is sincerely acknowledged. The authors would also like to acknowledge the Committee for Advanced Studies and Research (CASR), BUET.

\section*{Data availability}
The raw and processed data required to reproduce these findings cannot be shared at this time due to technical or time limitations.

\section*{Supplementary Information}
Additional electronic supplementary information (ESI) is available. See DOI: \href{https://doi.org/10.1016/j.jmrt.2021.06.026}{10.1016/j.jmrt.2021.06.026}


\bibliography{biblio} 


\end{document}